\documentclass[12pt,preprint]{aastex}

\def\dfrac#1#2{\displaystyle\frac{#1}{#2}}

\shorttitle{Effects of Relativistic Expansion on the Late-time Supernova
Light Curves}

\shortauthors{K.Iwamoto}

\begin{document}

\title{Effects of Relativistic Expansion on the Late-time Supernova
Light Curves}

\author{Koichi Iwamoto}

\affil{Department of Physics, College of Science and Technology, Nihon
University, \\ Tokyo 101-8308, Japan}
\email{iwamoto@phys.cst.nihon-u.ac.jp}

\begin{abstract}
The effects of relativistic expansion on the late-time supernova light
curves are investigated analytically, and a correction term to the
(quasi-)exponential decay is obtained by expanding the observed flux in
terms of \(\beta\), where \(\beta \) is the maximum velocity of the
ejecta divided by the speed of light \(c\).  It is shown that the
Doppler effect brightens the light curve owing to the delayed decay of
radioactive nuclei as well as to the Lorentz boosting of the photon
energies. The leading correction term is quadratic in \(\beta\), thus
being proportional to \(E_{\rm k}/(M_{\rm ej} c^2)\), where \(E_{\rm
k}\) and \(M_{\rm ej}\) are the kinetic energy of explosion and the
ejecta mass. It is also shown that the correction term evolves as a
quadratic function of time since the explosion.  The relativistic effect
is negligibly small at early phases, but becomes of considerable size at
late phases.  In particular, for supernove having a very large
energy(hypernova) or exploding in a jet-like or whatever non-spherical
geometry, \(^{56}\)Ni is likely to be boosted to higher velocities and
then we might see an appreciable change in flux. However, the actual
size of deviation from the (quasi-)exponential decay will be uncertain,
depending on other possible effects such as ionization freeze-out and
contributions from other energy sources that power the light curve.
\end{abstract}


\section{Introduction}

It has long been recognized that the radiative transfer in supernova(SN)
ejecta should be treated relativistically to account for the high
velocities achieved in its outermost layers.  However, it is only in the
last decade that the first numerical codes for radiative transfer that
take relativistic effects into account were developed and applied to
transfer problems in SN ejecta. At early times of explosion, the energy
of photons emitted from the photosphere of SN ejecta is greatly enhanced
by the Doppler effect.  As time goes on and the photosphere recedes
towards the center of ejecta, the degree of the enhancement decreases,
and eventually the resultant light curve(LC) is expected to follow an
(quasi-)exponential deposition curve due to radioactive decays. However,
there are several factors that cause the late-time LCs deviate from the
radioactive decay curve.  One possibility is the flattening of LCs owing
to the so-called ionization freeze-out effect, as first pointed out by
\citet{fra93} for the late-time LC of SN 1987A.  Another one is the
signatures of contributions from other energy sources such as possible
pulsar activity and circumstellar interaction as have been discussed by
previous works.

In this Letter, I would like to point out that late-time SN LCs may show
a deviation from radioactive decay curves as a result of the
relativistic Doppler effect.  In particular, for very energetic SNe or
what are called hypernovae(HNe), the relativistic effect would be very
important since the maximum velocity near the surface of ejecta reaches
a significant fraction of the speed of light.  It will be shown that the
Doppler effect makes a LC brighter due to the Lorentz boosting, but at
the same time the light-traveling-time effect and a pure-relativistic
effect delay the decay of radioactive nuclei, thus tending to make the
LC even brighter at late phases. The net result is determined by the sum
of these two effects.  Late-time LCs provide direct information to
determine the \(^{56}\)Ni masses ejected by SNe. Therefore, its precise
determination is of great importance for studies of SN explosion
mechanism and chemical evolution of galaxies.  In this Letter, we study
the relativistic effect on the late-time LCs and estimate the size of
the correction to the (quasi-)exponential decay for ordinary SNe and
HNe.  For simplicity, we use the approximation that energy input from
radioactivity is emitted as optical photons on the spot and carried away
from the SN ejecta free of absorption.

\section{Light Curves}

\begin{figure}
\plotone{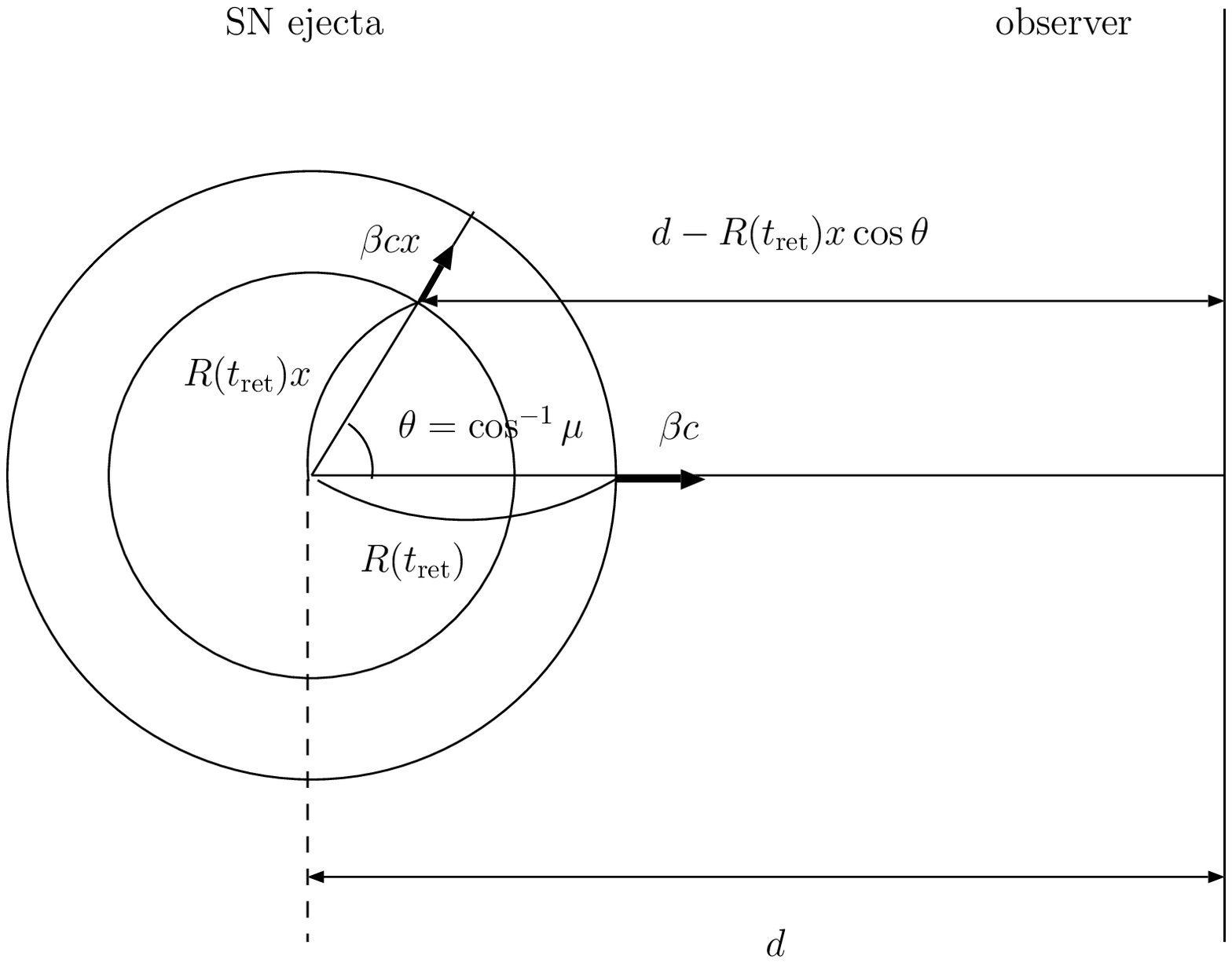} \vskip -7cm \caption{Schematic picture showing the
position of a volume element of SN ejecta at a retarded time \(t_{\rm
ret}\). \label{fig1}}
\end{figure}

We consider a spherically symmetric SN ejecta that expands homologously,
so that its radius is given by \( R(t) = R_{0} + \beta c t \) as a
function of time \(t\) after explosion, where \(R_0\) is the initial
radius, \(\beta c\) is the surface velocity, and \(c\) is the speed of
light.  Let us introduce the radial and azimuthal coordinates, \(x\) and
\(\mu\), as shown in Figure \ref{fig1}, to designate volume elements of
ejecta, such that \(x\) runs from 0 to 1 outwards and that \(\mu\) is
the cosine of angle \(\theta\) made between radial direction and the
line of sight.

The flux of photons received by an observer at time \(t\) is the sum of
the emissions from the volume element at a retarded time \(t_{\rm
ret}\), \( dV = 2 \pi R(t_{\rm ret})^3 x^2 dx d \mu \), where \(t_{\rm
ret}\) is defined by

\begin{eqnarray}
c(t-t_{\rm ret}) &=& d-R(t_{\rm ret}) x \cos \theta  \nonumber
\\
& \simeq & d -c \beta \mu x t_{\rm ret}
\label{eq:ret}
\end{eqnarray}

\noindent and \(d\) is the distance between the SN and the observer.  We
used the fact that the initial hydrodynamical time scale is short
compared to the elapsed time considered, \(R_0/\beta c << t \).  Solving
equation (\ref{eq:ret}) in terms of \(t_{\rm ret}\), we obtain

\begin{eqnarray}
t_{\rm ret} = \dfrac{t-d/c}{1-\beta \mu x}.
\end{eqnarray}

Let us define the frequency-integrated emissivity \(j\) as the total
energy of photons emitted per unit volume, unit time, and unit solid
angle.  Then, the emissivity in the rest frame in the direction toward
the observer \(j(x,\mu,t_{\rm ret})\) is related to that of the comoving
frame \(j_c\) such that
 
\begin{eqnarray}
j(x, \mu, t_{\rm ret}) = \dfrac{j_{\rm c}(x, \mu, t_{\rm ret, c})}
{\gamma_x^3(1-\beta \mu x)^3},
\end{eqnarray}

\noindent where \(\gamma_x = (1-\beta^2 x^2)^{-1/2}\) and \(t_{\rm
ret,c}=t_{\rm ret}/\gamma_x \) is the retarded time in the comoving
frame \citep{ryb79}. We assume that the radiative loss from the ejecta
is balanced by the energy input due to radioactive decay and that the
emissivity in the comoving frame is isotropic.  Then, \(j_c\) is written
as

\begin{eqnarray}
j_{\rm c}(x, \mu, t_{\rm ret, c}) = \frac{1}{4 \pi}
f \epsilon \dfrac{n_{\rm c}(x,t_{\rm ret})}{\tau},
\label{eq:jc}
\end{eqnarray}

\noindent where \(n_c\) is the number density of radioactive nuclei in
the comoving frame, \(\tau\) is its decay time, \(\epsilon\) is the
energy available per decay, and \(f\) is the deposition fraction.  The
total number of radioactive nuclei in a volume element \(\Delta V\),
\(\gamma_x n_{\rm c} \Delta V\), obeys the decay law given by

\begin{eqnarray}
\dfrac{\partial}{\partial t_{\rm ret, c}} (\gamma_x n_{\rm c} \Delta V)
= -\dfrac{\gamma_x n_{\rm c} \Delta V}{\tau},
\end{eqnarray}

\noindent
or

\begin{eqnarray}
\dfrac{\partial}{\partial t_{\rm ret}} (n_{\rm c} \Delta V)
= -\dfrac{n_{\rm c} \Delta V}{\gamma_x \tau},
\end{eqnarray}

\noindent
which has a solution


\begin{eqnarray}
n_{\rm c}(x,t_{\rm ret}) &=& \dfrac{\Delta V(0)}{\Delta V(t_{\rm ret})}
n_{\rm c}(x,0) e^{-\dfrac{t_{\rm ret}}{\gamma_x \tau}} =
\left(\dfrac{R_{0}}{R(t_{\rm ret})}\right)^3 n_{\rm c}(x,0)
e^{-\dfrac{t_{\rm ret}}{\gamma_x \tau}} \nonumber \\ &=&
\left(\dfrac{R_{0}}{R(t_{\rm ret})}\right)^3 \dfrac{n(x,0)}{\gamma_x}
e^{-\dfrac{t_{\rm ret}}{\gamma_x \tau}}.  \label{eq:nc}
\end{eqnarray}

Using equation (\ref{eq:nc}) in equation (\ref{eq:jc}), we obtain

\begin{eqnarray}
j_{\rm c}(x, \mu, t_{\rm ret, c}) 
= \left(\dfrac{R_{0}}{R(t_{\rm ret})}\right)^3 \frac{f \epsilon}{4 \pi}
\dfrac{n(x,0)}{\gamma_x \tau}e^{-\dfrac{t_{\rm ret}}{\gamma_x \tau}}.
\end{eqnarray}

In equating the energy emitted in a time \(d t_{\rm ret}\) by a volume
element at \(x, \mu\) and time \(t_{\rm ret}\), \(d V(x, \mu, t_{\rm
ret})\), with the corresponding energy passing through a normal area
\(\Delta S\) at the observer during time \(d t\), \( \Delta S dF(t)
dt\), we have

\begin{eqnarray}
dF(t) dt \Delta S = j(x, \mu, t_{\rm ret}) dt_{\rm ret} dV(x, \mu,
t_{\rm ret}) \Delta \Omega,
\end{eqnarray}

\noindent where \(dF(t)\) is the differential flux corresponding to the
volume element and \(\Delta \Omega\) is the solid angle subtended for
the area \(\Delta S\) by the volume element, and thus \(\Delta S = d^2
\Delta \Omega\). Adding up contributions by all the volume elements, we
have the observed flux at time \(t\),

\begin{eqnarray}
F(t) &=& \frac{1}{d^2}
\int \dfrac{\partial t_{\rm ret}}{\partial t} 
j(x,\mu, t_{\rm ret}) dV \nonumber \\
&=& 
\frac{1}{4 \pi d^2}
\dfrac{f N_{\rm tot}\epsilon}{\tau}
\int_0^1 dx 3 x^2 \tilde{n}(x) 
\times \frac{1}{2} \int_{-1}^{1} d \mu
\dfrac{1}{\gamma_x^4(1-\beta \mu x)^4}
\exp \left[-\dfrac{
\left(t-d/c\right)/\tau
}{\gamma_x (1-\beta \mu x)}
 \right].
\label{eq:flux1}
\end{eqnarray}

Here, \(\tilde{n}(x)=n(x,0)/\bar{n}\) is the dimensionless number
density of the radioactive nuclei and \(N_{\rm tot}=(4 \pi/3) R_{0}^3
\bar{n}\) is its total number at \(t=0\), where \(\bar{n}\) is the
average number density defined as

\begin{eqnarray}
\bar{n} \equiv \dfrac{\displaystyle\int_0^1 n(x,0) x^2 dx}
{\displaystyle\int_0^1 x^2 dx} = 3 \int_0^1 n(x,0) x^2 dx,
\end{eqnarray}

\noindent and thus \(\tilde{n}\) is normalized such that

\begin{eqnarray}
\int_0^1 3 x^2 \tilde{n}(x) dx = 1.
\end{eqnarray}

\noindent In deriving equation (\ref{eq:flux1}), it is assumed that
\(f\) is constant throughout the ejecta for simplicity.  This holds for
\(^{56}\)Co decay at sufficiently late times when the deposition is
primarily due to \(e^{+}\) decay, and even for general cases if only the
effect of a varying \(f\) is incorporated into \(\tilde{n}\). Using the
variable \(y \equiv \gamma_x^{-1} (1-\beta \mu x)^{-1}\) instead of
\(\mu\), the equation (\ref{eq:flux1}) is rewritten as

\begin{eqnarray}
F(t) = \frac{1}{4 \pi d^2}
\dfrac{f N_{\rm tot}\epsilon}{\tau}
\int_0^1 dx 3 x^2 \tilde{n}(x) 
f(\beta, x, X),
\label{eq:flux2b}
\end{eqnarray}

\noindent
where

\begin{eqnarray}
f(\beta, x, X) =
\frac{1}{2 \gamma_x \beta x}
\int_{a(\beta,x)}^{b(\beta,x)} y^2 e^{-Xy} dy,
\label{eq:flux2c}
\end{eqnarray}

\noindent
\(\displaystyle{X \equiv 
\frac{1}{\tau}\left(t-\dfrac{d}{c}\right), 
}\) \
and \(a(\beta,x), b(\beta,x)\) are given by

\begin{eqnarray}
a(\beta,x) = \left(
\frac{1-\beta x}{1+ \beta x}
\right)^{1/2}, \
b(\beta,x) = \left(
\frac{1+\beta x}{1- \beta x}
\right)^{1/2},
\end{eqnarray}

\noindent
respectively.
Note that \(b(\beta,x)=a(-\beta,x)\) so that \(f(\beta,x,X)\)
is an even function of \(\beta\).

The integration in equation (\ref{eq:flux2c}) can be readily done to
yield

\begin{eqnarray}
f(\beta, x, X) = \frac{g(b(\beta,x))-g(a(\beta,x))}
{2 \gamma_x \beta x},
\label{eq:FbxX}
\end{eqnarray}

\noindent
with

\begin{eqnarray}
g(y) = \int y^2 e^{-Xy} dy
=
-\frac{1}{X^3} e^{-Xy}
\left(
X^2 y^2 + 2X y +2
\right).
\label{eq:FbxX2}
\end{eqnarray}

Expanding \(g(a(\beta,x))\) and \(g(b(\beta,x))\)
in equation (\ref{eq:FbxX}) in terms of \(\beta\), 
we obtain 

\begin{eqnarray}
f(\beta, x, X) = g'(1) +
\left(\frac{g''(1)}{2}+\frac{g^{(3)}(1)}{6}\right) \beta^2 x^2 +
O(\beta^4) = e^{-X} \left[ 1+ \frac{X^2-7X+8}{6} \beta^2 x^2 +
O(\beta^4) \right]. \label{eq:expn}
\end{eqnarray}

Using equation (\ref{eq:expn}) in equation (\ref{eq:flux2b}) and
retaining terms up to the second order in \(\beta\), we get

\begin{eqnarray}
F(t) = \frac{1}{4 \pi d^2}
\dfrac{f N_{\rm tot}\epsilon}{\tau}
e^{-X}\left[
1+ \frac{X^2-7X+8}{6}
< \beta_x^2 >
\right],
\label{eq:flux3}
\end{eqnarray}

\noindent where \(<\beta_x^2>\) is the average of square velocity of
radioactive nuclei given by

\begin{eqnarray}
<\beta_x^2> = \int_0^1 dx
3 x^2 \tilde{n}(x) 
\beta^2 x^2 .
\label{eq:betax}
\end{eqnarray}

\section{Size of the Relativistic Correction}

The leading term in equation (\ref{eq:flux3}) corresponds to an
(quasi-)exponential decay of the flux that is expected for the
exponential decay of radioactive energy sources.  It is
quasi-exponential, because, in general, \(f\) changes with time. The
next term is a relativistic correction, which is of second order in
\(\beta\) and is generally very small. The correction is quadratic in
\(t' \equiv t-d/c\). It takes a value \((4/3) <\beta_x^2>\) at \(t'= 0\)
and has a minimum \(-(17/24) <\beta_x^2>\) at \( t' = 3.5 \tau\). At
late times when \(t' >> \tau\), the correction term becomes quite large.

To calculate \(<\beta_x^2>\), we need to know the density structure of
the ejecta. Here we assume a density profile 
 
\begin{eqnarray}
\rho(x)= \left\{
\begin{array}{lr}
\rho_c &  0 \leq x \leq x_c, \\
\rho_c \left( \dfrac{x}{x_c}\right)^{-p} & x_c \leq x \leq 1,
\end{array}
\right. 
\label{eq:density}
\end{eqnarray}

\noindent which is a good approximation to the actual SN ejecta and
greatly simplifies our analysis.  This profile is characterized by the
core radius \(x_c \) and the index of the power-law \(p\). For given
\(x_c\) and \(p\), \(\rho_c\) is determined to be

\begin{eqnarray}
\rho_c = 
\dfrac{p-3}{p x_c^3 -3 x_c^p}
\frac{3M_{\rm ej}}{4 \pi R^3}.
\nonumber
\end{eqnarray}

In addition, the maximum velocity achieved in SN ejecta, \(\beta\), is
given by

\begin{eqnarray}
\beta = \left(\frac{10 E_{\rm k}}{3 M_{\rm ej}
c^2} \cdot \frac{p-5}{p-3} \cdot \frac{p x_c^3-3 x_c^p}{p x_c^5-5 x_c^p}
\right)^{1/2},
\label{eq:beta}
\end{eqnarray}

\noindent where \(E_{\rm k}\) and \(M_{\rm ej}\) are the kinetic energy
of explosion and the ejecta mass, respectively.  Using the fact that
\(x_c \sim 0.1\) and \(p\) is as large as \(\sim 8-10\) for typical cases,
the equation (\ref{eq:beta}) reduces to

\begin{eqnarray}
\beta \simeq \left(\frac{10 E_{\rm k}}{3 M_{\rm ej} c^2}
\cdot \frac{p-5}{p-3}\right)^{1/2} x_c^{-1}, 
\label{eq:beta2}
\end{eqnarray}

\noindent which gives \(\beta \sim 0.33 (E_{\rm k}/10^{51} {\rm
erg})^{1/2} (M_{\rm ej}/M_{\odot})^{-1/2} (x_c/0.1)^{-1} \) for \(p =
8\).

If radioactive nucleus \(^{56}\)Ni is homogeneously mixed within the
ejecta and thus \(\tilde{n}(x)\) has the same profile as \(\rho(x)\), we
find from equation (\ref{eq:betax})

\begin{eqnarray}
< \beta_x^2 > = \frac{3}{5} \cdot \frac{p-3}{p-5} \cdot \frac{p x_c^5-5
x_c^p}{p x_c^3-3 x_c^p} \beta^2.
\label{eq:betax2}
\end{eqnarray}

From equations (\ref{eq:beta}) and (\ref{eq:betax2}), we obtain
\(<\beta_x^2 > = 2E_{\rm k}/(M_{\rm ej}c^2)\). Then, the correction term
in the bracket of equation (\ref{eq:flux3}) turns out to be

\begin{eqnarray}
\frac{X^2-7X+8}{6} < \beta_x^2 >
&=& 1.11 \times 10^{-3} \frac{X^2-7X+8}{6}
\left(\frac{E_{\rm k}}{10^{51} {\rm erg}}\right)
\left(\frac{M_{\rm ej}}{M_{\odot}}\right)^{-1}.
\label{eq:correction}
\end{eqnarray}

It is seen from equation (\ref{eq:correction}) that the size of the
relativistic correction depends only on \(M_{\rm ej}\) and \(E_{\rm
k}\), irrespective of the density profile of the ejecta in the
homogeneous-mixing case.

We use the fact that radioactivity is dominantly mediated by \(^{56}\)Co
decay of the decay chain \(^{56}$Ni$ \rightarrow ^{56}$Co$ \rightarrow
^{56}$Fe$\), and thus \(\tau = \tau_{\rm Co}-\tau_{\rm Ni} \equiv
111.3-8.8 = 102.5\) days and \(\epsilon = \epsilon_{\rm Co}\), the
energy available per decay of \(^{56}\)Co, effectively.  Then, it is
found that the correction term is only 0.16, 0.88, and 3.7 percent of
the leading term even at \(t' =\)2, 3, and 5 years, respectively, for
the case of Type Ia SN(SN Ia), which has a relatively large expansion
velocity because of its small mass \(M_{\rm ej} \sim M_{\odot}\) and
canonical explosion energy \(E_{\rm k} = 10^{51}\) erg.  For Type II
SNe(SNe II), the ejecta mass is much larger, e.g., \(M_{\rm ej} \sim 10
M_{\odot}\), so that the correction will be much smaller.  In the actual
SN ejecta, \(^{56}\)Ni is more centrally concentrated than the
homogeneous-mixing case now considered.  Thus, the above numbers are
likely to be an overestimate.  Therefore, it is concluded that, for
ordinary SNe, the relativistic effect would be very small compared to
other possible effects such as ionization freeze-out, circumstellar
interaction, and contributions from radioactivity of other than
\(^{56}\)Co.

\section{The Case of SN 1998bw}

\begin{figure}
\plotone{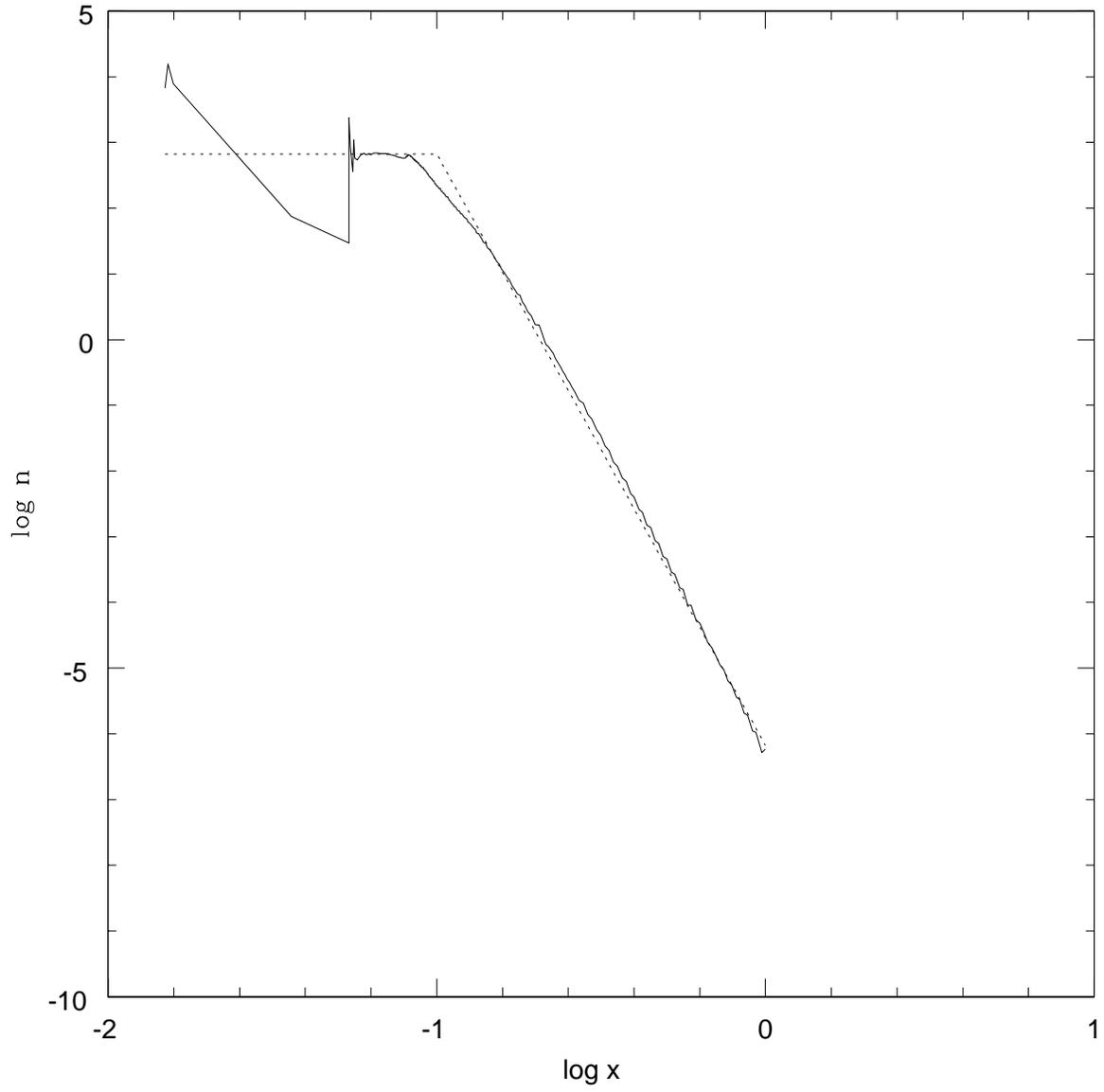} \caption{Normalized number density of \(^{56}\)Ni,
\(\tilde{n}\) as a function of \(x\), in the case of homogeneous mixing
for model CO138 of SN 1998bw.  The solid line is the result of
hydrodynamical calculation \citep{iwa98}, while the dashed line is an
analytic fit using equation (\ref{eq:density}) with \(p=10\) and \(x_c =
0.1\).  \label{fig2}}
\end{figure}

HNe have larger explosion energies than ordinary SNe by an order of
magnitude. Thus, we expect that the relativistic effect will be more
important for HNe.  As an example of HNe, we take a model for a
hypernova SN 1998bw, CO138 \citep{iwa98, nak01}, which has \(M_{\rm ej}
\sim 10 M_{\odot}\) and \(E_{\rm k} \sim (20-50) \times 10^{51}\) erg.
We found that the density distribution of CO138 is well reproduced by a
profile given by equation (\ref{eq:density}) with \(p = 10\) and \(x_c =
0.1\). Figure \ref{fig2} compares the analytic fit(dotted line) and the
result of hydrodynamical calculation (solid line) for \(\tilde{n}\) of
CO138 with \(E_{\rm k} = 30 \times 10^{51}\) erg \citep{iwa98} and with
homogeneous \(^{56}\)Ni distribution.  The value of \(p\) is in good
agreement with results by \citet{mat99} obtained for cases of compact SN
progenitors.

Then, from equation (\ref{eq:correction}), we find that the change of
the flux will be still relatively small at early times, but it grows
considerably at late times. It is only 1.8, 2.6, and 4.4 percent at \(t'
=3\) years, but becomes 7.4, 11.1, and 18.5 percent at \(t' = 5\) years
for cases of \(E_{\rm k} =\) 20, 30, and \(50 \times 10^{51}\) erg,
respectively.  If the explosion occurs in a jet-like \citep{nag00} or
whatever non-spherical geometry, \(^{56}\)Ni is likely to be boosted to
higher velocities and then we might see an appreciable change in
flux. In late-time spectra of SN 1998bw, it was observed that iron has a
larger velocity than oxygen, indicating that such a non-spherical
\(^{56}\)Ni distribution may be realized in the ejecta of SN 1998bw
\citep{maz01}.
 
Even for spherically symmetric models, it is reported that the ejecta of
HNe may become trans-relativistic with its outer envelope having a mean
velocity \(\gamma = (1-\beta^2)^{-1/2} \sim 2\) (\(\beta \sim
\sqrt{3}/2\)) and a flatter density gradient of \( p \sim 4 \)
\citep{tan01}.  Therefore, to make a detailed comparison between the
observed late-time LCs and those predicted by non-spherical HN models,
it is recommended that we calculate model LCs by taking the relativistic
effect into account.  Note that for large values of \(\beta\) the
expansion series in terms of \(\beta\) (equation \ref{eq:expn}) may not
converge well. However, the analytic solution with equations
(\ref{eq:FbxX}) and (\ref{eq:FbxX2}) is exact, and thus we can use it to
study such cases with large \(\beta\).

\bigskip

The author would like to thank Drs. Shiomi Kumagai, Takayoshi Nakamura,
and Professor Emeritus Masatomo Sato for useful discussion.  This work
has been supported in part by the Grant-in-Aid for Scientific Research
(12740122, 12047208) of The Ministry of Education, Culture, Sports,
Science, and Technology(MEXT) in Japan.

\clearpage

\clearpage

\begin{thebibliography}{}

\bibitem[Fransson \& Kozma(1993)]{fra93} Fransson, C., and Kozma,
C. 1993, \apj, 408, L25

\bibitem[Iwamoto et al.(1998)]{iwa98} Iwamoto, K. et al. 1998, \nat,
395, 672

\bibitem[Matzner \& McKee(1999)]{mat99} Matzner, C. D., and McKee,
C. F. 1999, \apj, 510, 379

\bibitem[Mazzali et al.(2001)]{maz01} Mazzali, P. A., Nomoto, K., Patat,
F., and Maeda, K. 2001, \apj, 559, 1047

\bibitem[Nagataki(2000)]{nag00} Nagataki, S. 2000, \apjs, 127, 141

\bibitem[Nakamura et al.(2001)]{nak01} Nakamura, T., Mazzali, P. A.,
Nomoto, K., and Iwamoto, K.  2001, \apj, 550, 991

\bibitem[e.g.,Rybicki \& Lightman(1979)]{ryb79} Rybicki,G.B., \&
Lightman,A.P. 1979, Radiative Processes in Astrophysics (John Wiley \&
Sons)

\bibitem[Tan, Matzner, \& McKee(2001)]{tan01} Tan, J. C., Matzner,
C. D., and McKee, C. F. 2001, \apj, 551, 946

\end{thebibliography}
\end{document}